\documentstyle[prl,aps,psfig,graphicx,epsf,multicol]{revtex}

\def\mbf(#1){\mbox{\boldmath $#1$}} 

\begin{document}
\title{Nature of insulating state in NaV$_{2}$O$_{5}$ above charge-ordering
transition: a cluster DMFT study}
\author{V. V. Mazurenko$^{1,2}$, A. I. Lichtenstein$^{1}$, 
M. I.\ Katsnelson$^{1,2}$, I. Dasgupta$^{3}$, T. Saha-Dasgupta$^{4}$ and 
V. I. Anisimov$^{2}$}
\address{$^{1}$University of Nijmegen, NL-6525 ED Nijmegen, The Netherlands\\
$^{2}$Institute of Metal Physics, Ekaterinburg, Russia\\ 
$^{3}$ Department of Physics, Indian Institute of Technology,
Bombay, Mumbai 400 076; \\ $^{4}$ S.N. Bose National Center for Basic 
Sciences, JD Block, Salt Lake City, Calcutta 700 091}
\date{\today}
\maketitle

\begin{abstract}
The nature of insulating state driven by electronic correlations in
the quarter-filled ladder compound 
$\alpha'$NaV$_{2}$O$_{5}$ is investigated within a cluster dynamical
mean-field approach.
An extended Hubbard model with first-principle tight-binding parameters
have been used. It is shown that the insulating state in the
charge-disordered phase of this compound
is formed due to the transfer of spectral density and dynamical charge
fluctuations where for the latter, the role of inter-site Coulomb 
interaction is found to be of crucial importance.
\end{abstract}

\begin{multicols}{2}[]
The ladder compound NaV$_{2}$O$_{5}$  has been a subject of great interest since last five
years \cite{japan,weber,optics,HF,khomskii,XRay,NMR,Sa,yaresko,johnston}. It
exhibits a remarkable phase transition at T$_c$ =34 K, now identified
as charge-ordering of {\it zigzag} type \cite{weber,HF,khomskii,XRay,NMR}%
. Both charge-ordered phase and charge-disordered one are insulating with an
energy gap of the order of 0.8-1 $eV$ \cite{optics}. The presence of the gap in
the ordered phase is not surprising and it was reproduced successfully, for
example, in the recent local density approximation(LDA)+U calculations \cite{yaresko}. The properties of the
disordered phase are much more difficult to understand. In contrast to
the iso-structural ladder compound, CaV$_{2}$O$_{5}$, NaV$_{2}$O$_{5}$ has
quarter-filled band \cite{weber} \ rather than half-filled one and cannot be
considered as a {\it standard} Mott insulator \cite{mott}. It has been
proposed in Ref. \cite{khomskii} that a large value of the
transverse hopping parameter in the
ladder which splits the band into sub-bands of the bonding and anti-bonding
states \cite{weber} could be responsible for the Mott insulator behaviour in
NaV$_{2}$O$_{5}$  in the presence of 
 strong Coulomb interactions. However
it is not known whether this mechanism is adequate for the realistic values of the parameters characterising 
the single-particle electronic
structure and the electron-electron correlations. 

In this letter we investigate the correlation effects in NaV$_{2}$O$_{5}$ 
taking into account non-local dynamical charge fluctuations
on the rung for this two-leg ladder compound. This allows us
to understand the origin of the insulating states above T$_c$ and
to estimate the relative importance of various physical mechanisms
responsible for the gap formation.

The crystal structure of NaV$_{2}$O$_{5}$  projected in the xy plane is schematically shown in Fig.1. The results
of the X-ray \cite{XRay}, NMR \cite{NMR}, and optical \cite{optics}
experiments as well as the Hartree-Fock calculations \cite{HF} support the
{\it zigzag} charge-ordering state for low temperatures (see Fig.1). In this
state one has approximately one $d$ electron per rung of the vanadium ladder. We
start with LDA+U \cite{LDA+U} calculations of the ordered states but in
contrast to the previous work \cite{yaresko} we considered several different types of the
charge ordering. This gives us an opportunity to estimate the  on-site and
inter-site Coulomb interactions U and V  respectively
which in turn were used to parametrise the model Hamiltonians to be used for 
the calculations taking into account the dynamical correlation effects.

\begin{figure}[t]
\begin{center}
\begin{minipage}[t]{8cm}
\epsfxsize=8cm
\epsfbox{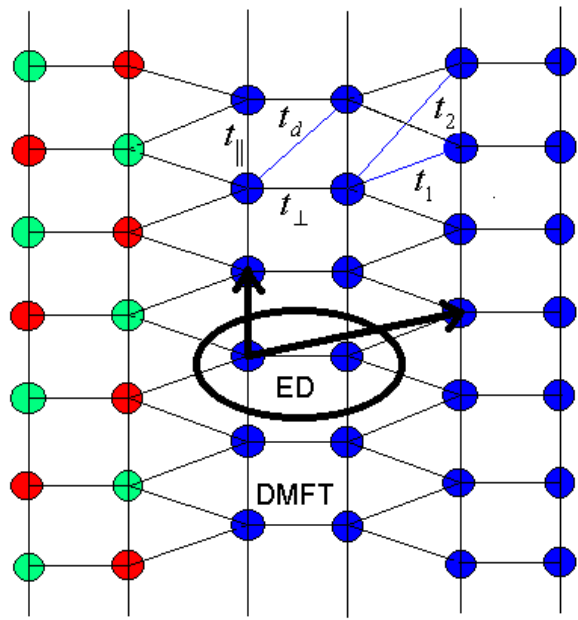}
\caption {Schematic representation of the crystal structure of the vanadium 
layers in NaV$_{2}$O$_{5}$ and the hopping matrix elements.
The vanadium ions are denoted by filled circles. The ellipse shows the 
cluster which plays the role of an effective impurity in the DMFT calculations. 
A {\it zig-zag} charge ordering of the V$^{4+}$ and V$^{5+}$ ions,
obtained from our LDA+U calculations as 
a ground state is shown on the left ladder. Bold arrows are the translation vectors.}
\end{minipage}
\end{center}
\label{struc}
\end{figure}

By mapping the results of the LDA+U calculations for different types of
charge ordering on the results of model calculations with on-site (U)
and inter-site (V) Coulomb interaction parameters, we obtained the following
values: U=2.8 eV and V=0.17 eV.

It is natural to assume that the tendency to keep the number of $d$
electrons per rung close to unity also takes place above the transition
temperature leading to strong short-range order and well-developed dynamical
charge fluctuations. This is confirmed by the temperature dependences
of the spin gap and the entropy measurements\cite{johnston}.  
Usual LDA as well as the mean-field theories like
Hartree-Fock or LDA+U methods are insufficient to take into account these
essential many-body processes. The simplest reliable way to consider such
short-ranged correlation effects is the use of the dynamical mean-field theory (DMFT) 
\cite{review} which can be combined with realistic LDA band structure
calculations (LDA+DMFT) \cite{anisimov,LK,voll}. The DMFT maps the
initial many-body problem for a crystal onto a self-consistent quantum-impurity
problem. To consider the phenomena such as charge ordering
or fluctuations, the inter-site correlations are of crucial importance, therefore we
need to use a cluster generalisation of the DMFT 
method \cite{cluster1,cluster2,cluster3}. The most reasonable 
choice of the cluster in our case is a pair of vanadium atom at the rung
(see Fig.1).  The unique geometry of the ladder compounds 
makes the choice of proper cluster simpler and the
free cluster consideration is the most natural choice\cite{cluster2,cluster3}.

\begin{figure}[t]
\begin{center}
\begin{minipage}[t]{8cm}
\epsfxsize=8cm
\epsfbox{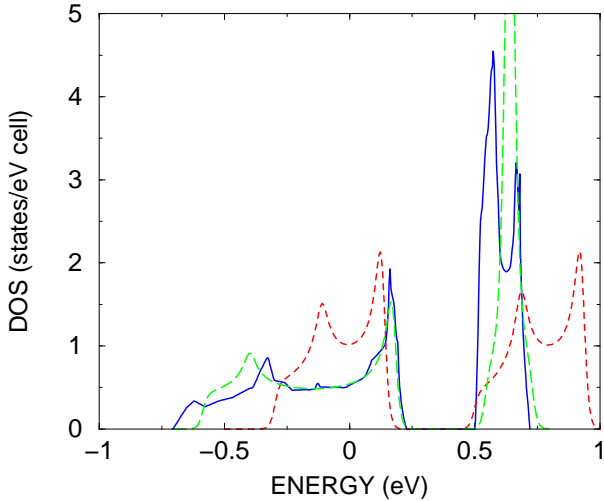}
\caption {Bare densities of states (DOS). Solid lines correspond to LDA DOS. 
The dashed and long dashed lines represent the first-principle tight-binding 
parameterisations without and with t$_{d}$ hopping respectively. 
}
\end{minipage}
\end{center}
\label{dos}
\end{figure}

We use the extended Hubbard model for two-leg ladder:

\[
H=\sum_{ij}t_{ij}c_{i\sigma }^{+}c_{j\sigma }+\sum_{i}U_{i}n_{i\uparrow
}n_{i\downarrow }+\sum_{i<j,\sigma }V_{ij}n_{i\sigma }n_{j\sigma } 
\]
where $t_{ij}$-s are the effective hoppings, $U_{i}$ and $V_{ij}$ are local
and inter-site Coulomb interactions respectively, $n_{i\sigma }=$ $c_{i\sigma
}^{+}c_{i\sigma }$. The  proper choice of hopping parameters
is not simple, and the most widely used set  
( $t_{\perp }=0.38$ eV, $t_{\parallel }=0.18$
eV, $t_{1}=0.012$ eV, $t_{2}=0.03$ eV ) was obtained by the fitting
to the LDA bands \cite{weber}. Recently a rigorous procedure
of {\it massive downfolding} of LDA bands to a few-band description and the subsequent Fourier transformation
of the resulting Hamiltonian from the reciprocal to direct space
to extract the single-electron parameters has been
developed within the framework of linear-muffin-tin-orbital (LMTO)
description \cite{downfolding}. This method applied to NaV$_{2}$O$_{5}$
gave the following set of hopping parameters:
t$_{\bot}$=0.398 eV, t$_{\parallel}$=0.084 eV, t$_{1}$=0.025 eV, t$_{2}$=0.022 eV
which is rather close to the standard one presented above,
but in addition the {\it diagonal} (Fig. 1) hopping parameter t$_{d}$=0.083 eV
is appreciable and was not considered before. 
We have found that these diagonal hopping processes are very
important. As can be see in Fig. 2, the inclusion of diagonal
hopping t$_{d}$ in the single-electron part of the model Hamiltonian
provides a much better agreement of the bare DOS with
the LDA DOS. It also results in the insulating
state of the charge-disordered systems for the realistic values of
U and V obtained from LDA+U calculations, while the
other TB parametrisation, excluding the diagonal
hopping, results into the metallic state for the same
values of U and V (see Fig. 3).

The crystal Green function matrix in LDA+DMFT approach can be
written as 
\[
{\bf G}\left( {\bf k,}i\omega \right) =\left[ (i\omega +\mu )\ast {\bf 1}-%
{\bf h}\left( {\bf k}\right) -{\bf \Sigma }\left( i\omega \right) \right]
^{-1} 
\]
where $h_{\alpha \beta }\left( {\bf k}\right) $ is the effective hopping
matrix, $\Sigma _{\alpha \beta }\left( i\omega \right) $ is the self-energy
matrix of the two-site super-cell dimension which is assumed to be local, i.e. ${\bf k}$-independent,
and $\mu $ is the chemical
potential.

In the cluster version of the DMFT scheme\cite{review,cluster2}, one can write the
matrix equation for a bath Green function matrix ${\cal G}$ which describe
an effective interactions with the rest of the crystal: 
\begin{equation}
{\cal G}^{-1}\left( i\omega \right) ={\bf G}^{-1}\left( i\omega \right) +%
{\bf \Sigma }\left( i\omega \right)   \label{cavity}
\end{equation}
where the local cluster Green function matrix is equal to $G_{\alpha \beta
}\left( i\omega \right) =$ $\sum\limits_{{\bf k}}G_{\alpha \beta }\left( 
{\bf k,}i\omega \right) $ , and the summation runs over the Brillouin zone of
the lattice.

We used the exact diagonalisation (ED) scheme to solve the cluster DMFT problem.
In this case the lattice Hamiltonian 
is mapped onto finite cluster impurity model:
\begin{eqnarray}
H_{imp} && = \sum_{i,j,\sigma }T_{ij}c_{i\sigma }^{+}c_{j\sigma
}+\sum_{i}U_{i}n_{i\uparrow }n_{i\downarrow }+\sum_{i<j,\sigma
}V_{ij}n_{i\sigma }n_{i\sigma }  \label{imp} \\
+\sum_{k,i,j,\sigma }&&E_{ij\sigma }(k)a_{ki\sigma }^{+}a_{kj\sigma }
+\sum_{k,i,j,\sigma }\Gamma _{ij\sigma }(k)\left( a_{ki\sigma
}^{+}c_{j\sigma }+h.c.\right)   \nonumber
\end{eqnarray}
where $T_{ij}$ is the hopping parameters inside the cluster (for our two site
rung-cluster this is only $t_{\perp }$), $E_{ij\sigma }(k)$, and $\Gamma
_{ij\sigma }(k)$ are effective energies and hybridisation matrix for
finite-chain bath orbitals $k=1,...,n_{b}$. 

For the iterative solution of the effective impurity model Eq.(\ref{imp}) we use the
Lanczos version of ED method\cite{review}.
The orbital energy matrix
for the conduction band $E_{ij\sigma }(k)$, and the corresponding
hybridisation elements $\Gamma _{ij\sigma }(k),$ are the effective
parameters which reproduce the bath Green function:

\begin{equation}
{\cal G}^{-1}(i\omega _{n})=(i\omega _{n}+\mu )\ast {\bf 1}-{\bf T}-\sum_{k}%
{\bf \Gamma }_{k}[i\omega _{n}-{\bf E}_{k}]^{-1}{\bf \Gamma }_{k}^{+}
\end{equation}
In the paramagnetic case we can transform ${\cal G}$, ${\bf T}$, ${\bf %
\Gamma }_{k}$, and ${\bf E}_{k}$ matrices of the dimension $2\times 2$ for
our cluster to the diagonal {\it bonding-antibonding} basis $\lambda =\{b,a\}$%
. In this case $T_{\lambda }=\{-t_{\perp },t_{\perp }\}$ and ${\cal G}%
_{\lambda }=\{{\cal G}_{b},{\cal G}_{a}\}$, where ${\cal G}_{b,a}=({\cal G}%
_{11}\pm {\cal G}_{12})/2$.
We used 10 bath orbitals for our two-site cluster.

The parameters $\{E_{k\lambda \sigma },\Gamma _{k\lambda \sigma }\}$ are now
fitted to reproduce the bath Green function ${\cal G}_{\lambda \sigma
}(i\omega _{\nu })$ for bonding and anti-bonding states independently. The
next step is the solution of the cluster-impurity problem (Eq. (\ref{imp})) to
get the cluster self-energies $\Sigma _{ij\sigma }(i\omega _{\nu })$ which
are required for the next DMFT-iterations. After solving the effective
cluster problem using Lanczos algorithm, the local Green's function matrix $%
G_{ij\sigma }(i\omega _{\nu })$ was determined and the cluster self-energy $%
\Sigma _{ij\sigma }(i\omega _{\nu })$ was obtained within the 
cluster-impurity scheme (Eq. (\ref{cavity})).

The phase digram obtained in our DMFT calculations is
shown in Fig. 3. One can see that for large
enough $U$ and realistic values of $V$ the disordered state turns out to be insulating. This
arises from the two physically different mechanisms. The first one is the
spectral density transfer which can effectively change the
quarter-filled system into the half-filled one. For large enough $U$ the
energy band splits into Hubbard sub-bands with the average spectral
weight 1/2 for each of them and therefore the system appears to be insulator
for half-filling instead of complete filling \cite{hubbard}. 
In this {\it large-U} limit the Hubbard model reduces to so-called t-J-V
model and can explain insulating properties and optical spectra of
NaV$_{2}$O$_{5}$ \cite{cuoco}.
Our calculations do show a formation of the lower and upper Hubbard bands
which leads to the spectral density transfer
and moves the Fermi energy into the pseudo-gap between
bonding and anti-bonding states in the lower Hubbard band
 as it was proposed qualitatively in Ref. \cite
{khomskii}. However, the gap which can be obtained due to this
mechanism is very small. To increase its value
(consistent with the experiment\cite{optics}) the inclusion of the
inter-site Coulomb repulsion (V)
appears to be important (Fig. 4). Our calculations show that
the broad enough gap arises  with the increase of $V$ suddenly, as a result of
a first-order phase transition.
With V=0 the insulator is stable  only for U $\approx$ 4 eV and above, while already
small value of V=0.1 eV decreases the critical U value below 3 eV.
The results of LDA+U calculations (U=2.8 eV and V=0.17 eV) gave a point
in the phase diagram (the cross on Fig. 3) which is well above the metal-insulator
transition line.
It is important to stress that if one carries out the calculation in usual
single-site DMFT
calculation instead of the cluster DMFT an adequate description of 
the electronic structure is not obtained for the same
values of the parameters, and the insulating states appears only
for U $>$ 12 eV!
This demonstrates the crucial importance of the charge fluctuations on the rung for
the formation of insulating state in NaV$_{2}$O$_{5}$ with the realistic U-values.
We also analysed the structure of the ground state in the effective cluster model
(Eq.(\ref{imp})) and found that for our LDA+U Coulomb parameters about 70$\%$ 
of the ground state eigenfunction
corresponds to the configurations with one electron per rung and 20$\%$ 
related to empty rung states. This means that the configurations with 
the empty rungs in NaV$_{2}$O$_{5}$ has also appreciable weight.

\begin{figure}[t]
\begin{center}
\begin{minipage}[t]{8cm}
\epsfxsize=8cm
\epsfbox{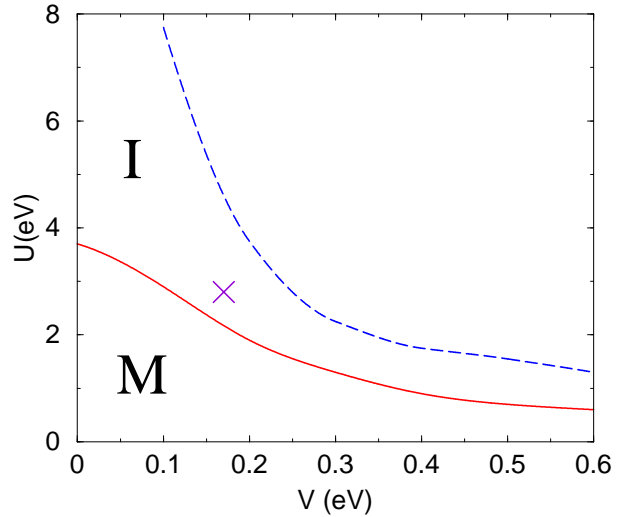}
\caption{Calculated phase diagram for the model with hopping parameter
from Ref. {\protect\cite{weber}} (dashed line) and the parameters obtained from our
TB-LMTO calculation (solid line). The lines
demark the metallic and insulating phases in two calculations. The cross
corresponds the values of $U$ and $V$ obtained from our LDA+U
calculation.}
\end{minipage}
\end{center}
\label{phase}
\end{figure}

We conclude that the insulating state of NaV$_{2}$O$_{5}$ 
above $T$ =34 $K$ is characterised by strong dynamical charge fluctuations.
In this respect the situation is close to the half-filled Hubbard chain or 
an {\it pseudospin-liquid} like phase: 
the tendency for the formation of the state with one electron per rung is
similar to 
the singlet state formation in strongly frustrated
spin systems
and can be described in the limit of large $U$ and $V$ by 
anisotropic Heisenberg-like model \cite{IK,khomskii,Sa,new}. 
It is important to stress that experimentally the spin gap exists also above
T$_c$ \cite{johnston}. However, this description is
only qualitative.

\begin{figure}[t]
\begin{center}
\begin{minipage}[t]{8cm}
\epsfxsize=8cm
\epsfbox{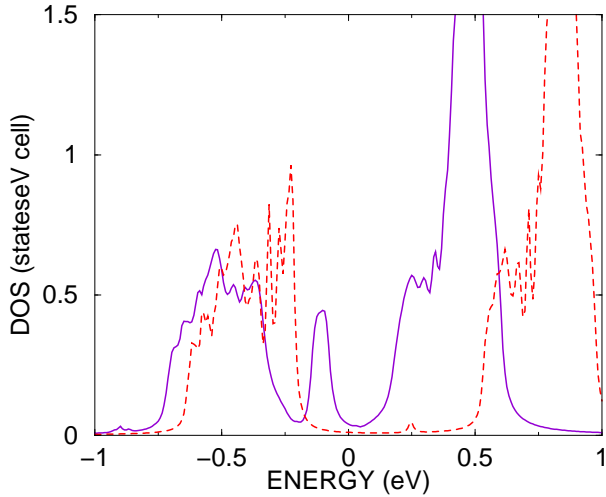}
\caption {The renormalised densities of state (only the lower Hubbard bands)
obtained in our DMFT calculation with 
parameter $U$=2.8 eV and $V$=0.17 eV(solid line); $U$=2.8 eV and $V$=0.5 eV(dashed line).}
\end{minipage}
\end{center}
\label{dosDMFT}
\end{figure}  

Our
calculations demonstrate that this compound is situated near the boundary
between metal and insulator states and should be described by
itinerant-electron models rather than by localised-electron ones. Since the
metal-insulator transition in the model under consideration turns out to be
strongly first-order type, the gap is not small.
The value of the gap depends strongly on the inter-site Coulomb
repulsion parameter V, as is demonstrated in Fig. 4. The insulating properties cannot be described
correctly neglecting the charge fluctuations on the rung (too small values of $%
V $ or using single-site DMFT approach instead of the cluster one).

To conclude, we present for the first time results on the cluster DMFT calculations
for the insulating phase of NaV$_{2}$O$_{5}$ above charge-ordering transition
using the extended Hubbard model with the first-principle tight-binding
parameters. The non-local charge fluctuations and inter-site Coulomb interaction
is of crucial importance for the formation of insulating state.

This work is supported by the Nederlandse Organisatie voor Wetenschappelijk
Onderzoek (NWO), project 047-008-16 and 047-008-012 and Russian Foundation for
Basic Research grant RFFI-01002-17063.

\end{multicols}

\begin{references}
\bibitem{japan}  M. Isobe and Y. Ueda, J. Phys. Soc. Japan {\bf 65,} 1178
(1996).

\bibitem{weber}  H. Smolinski {\it et al}, Phys. Rev. Lett. {\bf 80,} 5164
(1998).

\bibitem{optics}  C. Presura, D. van der Marel, A. Damascelli, and R. K.
Kremer, Phys. Rev. B {\bf 61,} 15762 (2000).

\bibitem{HF}  H. Seo and H. Fukuyama, J. Phys. Soc. Japan {\bf 67,} 2602
(1998).

\bibitem{khomskii}  M. V. Mostovoy and D. I. Khomskii, Solid State Commun. 
{\bf 113, }159 (2000).

\bibitem{XRay}  H. Nakao {\it et al}, cond-mat/0003129.

\bibitem{NMR}  T. Ohama et al, cond-mat/0003141.

\bibitem{Sa}  D. Sa and C. Gros, cond-mat/0004025.

\bibitem{yaresko}  A. N. Yaresko, V. N. Antonov, H. Eschrig, P.Thalmeier,
and P. Fulde, Phys. Rev. B {\bf 62, }15538 (2000).

\bibitem{johnston} D. C. Johnston {\it et al}, 
 Phys. Rev. B {\bf 61, } 9558 (2000).

\bibitem{mott}  N. F. Mott, {\it Metal-Insulator Transitions }(Taylor and
Francis, London, 1974).

\bibitem{LDA+U}  V. I. Anisimov, F. Aryasetiawan, and A. I. Lichtenstein, J.
Phys.: Condens. Matter {\bf 9,} 767 (1997).

\bibitem{review} For a review see:
A. Georges, G. Kotliar, W. Krauth, and M. J. Rozenberg, %
Rev. Mod. Phys. {\bf 68}, 13 (1996).

\bibitem{anisimov}  V. I. Anisimov, A. I. Poteryaev, M. A. Korotin, A. O.
Anokhin, and G. Kotliar, J. Phys.: Condens.\ Matter {\bf 9, }7359 (1997).

\bibitem{LK}  A. I. Lichtenstein and M. I. Katsnelson, Phys. Rev. B {\bf 57, 
}6884 (1998); M. I. Katsnelson and A. I. Lichtenstein, J. Phys.: Condens.\
Matter {\bf 11, }1037 (1999); Phys. Rev. B {\bf 61, }8906 (2000).

\bibitem{voll}  I. A. Nekrasov {\it et al}, Eur. Phys. J. B {\bf 18,} 55
(2000); M. B. Zoelfl {\it et al}, Phys. Rev. B {\bf 61, }12810 (2000).

\bibitem{cluster1}  M. H. Hettler {\it et al}, Phys. Rev. B {\bf 58, }7475
(1998); Phys. Rev. B {\bf 61, }12739 (2000).

\bibitem{cluster2} A. I. Lichtenstein and M. I.
Katsnelson, Phys. Rev. B {\bf 62,} R9283 (2000).

\bibitem{cluster3} G. Kotliar, S. Y. Savrasov,
and G. Palsson, cond-mat/0010328.

\bibitem{downfolding} O. K. Andersen and T. Saha-Dasgupta, 
Phys. Rev. B {\bf 62, } R16219  (2000).

\bibitem{hubbard}  J. Hubbard, Proc. Roy. Soc. (London) A {\bf 276,} 238
(1963); A {\bf 281,} 401 (1964).

\bibitem{cuoco}  M. Cuoco, P. Horsch, and F. Mack,
Phys. Rev. B {\bf 60, } R8438 (1999).

\bibitem{IK}  V. Yu. Irkhin and M. I. Katsnelson, JETP Lett. {\bf 49,} 576
(1989); Phys. Lett. A {\bf 150,} 47 (1990).

\bibitem{new} A. Langari, M. A. Martin-Delgado, and P. Thalmeier, cond-mat/0102007
\end{references}
\end{document}